\documentclass[aps,pra,twocolumn,showpacs,superscriptaddress,10pt]{revtex4-1}

\usepackage{amsmath}
\usepackage{amssymb}
\usepackage{graphicx}
\usepackage[colorlinks=true,linkcolor=blue,urlcolor=cyan,citecolor=red,anchorcolor=blue]{hyperref}

\begin{document}

\title{Delta-Kick Collimation of Heteronuclear Feshbach Molecules}

\author{T. Estrampes}
\affiliation{Leibniz Universit\"at Hannover, Institut f\"ur Quantenoptik, Welfengarten 1, 30167 Hannover, Germany}
\affiliation{Universit\'e Paris-Saclay, CNRS, Institut des Sciences Mol\'eculaires d'Orsay, 91405 Orsay, France}
\author{J. P. D'Incao}
\affiliation{JILA, NIST, and the Department of Physics,
University of Colorado, Boulder, CO 80309, USA}	
\affiliation{Department of Physics, University of Massachusetts Boston, Boston, MA 02125, USA}	
\author{J. R. Williams}
\affiliation{Jet Propulsion Laboratory, California Institute of Technology, Pasadena, CA, USA}
\author{T. A. Schulze}
\affiliation{Leibniz Universit\"at Hannover, Institut f\"ur Quantenoptik, Welfengarten 1, 30167 Hannover, Germany}
\author{E. M. Rasel}
\affiliation{Leibniz Universit\"at Hannover, Institut f\"ur Quantenoptik, Welfengarten 1, 30167 Hannover, Germany}
\author{E. Charron}
\affiliation{Universit\'e Paris-Saclay, CNRS, Institut des Sciences Mol\'eculaires d'Orsay, 91405 Orsay, France}
\author{N. Gaaloul}
\affiliation{Leibniz Universit\"at Hannover, Institut f\"ur Quantenoptik, Welfengarten 1, 30167 Hannover, Germany}

\begin{abstract}
We present a theoretical study of delta-kick collimation (DKC) applied to heteronuclear Feshbach molecules, focusing on both condensed and thermal ensembles across various interaction and temperature regimes. We demonstrate that DKC enables significant reductions in molecular cloud expansion energies and beam divergence, achieving expansion energies in the picokelvin range, comparable to state-of-the-art results obtained experimentally with atoms. Furthermore, we show that vibrational and translational motions remain strongly decoupled throughout the process, ensuring molecular stability during the delta-kick. This work paves the way for advanced experimental sequences involving degenerate ground state molecules, light-pulse molecular interferometry, and applications of dual-species precision measurements, such as testing the universality of free fall.
\end{abstract}

\maketitle

\section{Introduction and Motivation}
\label{sec:intro}

Magnetic Feshbach resonances play an important role in the study of ultracold gases, providing a versatile tool for a wide range of applications\;\cite{RevModPhys.82.1225}. By applying a bias magnetic field, these resonances enable precise tuning of the interparticle interactions, universally characterized by the atomic scattering length $a$, spanning regimes from strongly attractive or repulsive to non-interacting. Moreover, Feshbach resonances allow the adiabatic formation and dissociation of extremely weakly bound diatomic molecules. These Feshbach molecules\;\cite{FESHBACH1958357, TIMMERMANS1999199} are instrumental in the study of various physical phenomena, such as the BEC-BCS crossover\;\cite{greiner2003Nature, cubizolles2003prl, regal2004prla, jochim2003Science, zwierlien2004prl, strecker2003prl, regal2004prlb, zwierlein2004prl, chin2004Science, zwierlein2005Nature}, the creation of ultracold polar molecules\;\cite{ospelkaus2008NatPhys, ni2008Science, zirbel2008prl, wu2012prl, heo2012pra, tung2013pra, repp2013pra, koppinger2014pra, wang2013pra, takekoshi2012pra, deh2010pra}, and the study of universal few-body phenomena\;\cite{braaten2006PR,dincao2018JPB}. They also enable the controlled creation of entangled states\;\cite{greiner2005prl, poulsen2001pra, kherunssyan2002pra, kheruntsyan2005pra, yurovsky2003pra, savage2007prl, kheruntsyan2005prl, zhao2007pra, davis2008pra, gneiting2008prl, gneiring2010pra}, and serve as a platform for testing variations of fundamental constants\;\cite{chin2006prl, chin2009njp, borschevsky2011pra, gacesa2014jms}. However, in most cases, the temperature at which the Feshbach molecule is created is a critical factor that can limit its applicability.
In atom interferometry, reducing the free-expansion rate (i.e., energy) of atomic ensembles and Bose-Einstein condensates (BECs)\;\cite{WiemanCornell1995, Ketterle1995} is a crucial step allowing for the precise control of the system's dynamics and minimization of systematic errors\;\cite{Schlippert2020}. Various collimation techniques have been developed to address this issue\;\cite{PhysRevLett.78.2088, Chu86OL,  Wang2023, Albers2022, PhysRevResearch.6.013139}. The most successful results to date relied on the delta-kick collimation (DKC) technique in rubidium systems\;\cite{Kovachy2015, PhysRevLett.127.100401, Gaaloul2022}.  Recently, the control of interactions using a Feshbach resonance coupled with a matter-wave lensing protocol has allowed the collimation of $^{39}$K BECs to sub-nK energies\;\cite{Herbst24} in the presence of gravity, with the prospect of reaching double-digit pK energies by incorporating a DKC stage\;\cite{PhysRevLett.78.2088}. It is natural to seek to extend the advantages of operating at ultra-low energies to dual-species systems. However, extending these techniques to the simultaneous collimation of dual-species gases presents significant challenges, primarily due to mismatches in trap frequencies and different in-trap dynamics of the two species\;\cite{Meister2023}. Similarly, despite the remarkable progress in collimating ultracold atomic ensembles using DKC, its extension to molecular systems, particularly heteronuclear Feshbach diatomic molecules, remains largely unexplored. The additional complexities arising from coupled molecular degrees of freedom, such as vibrational and translational motions, as well as the potential for species-specific dynamics, present significant challenges to the straightforward application of existing techniques. Addressing these challenges is critical for advancing the control of ultracold molecular ensembles and unlocking their potential for precision measurements and quantum technologies.

In this work, we aim to bridge this gap by revisiting the DKC technique to extend its application range for Feshbach molecular ensembles in both thermal and condensed regimes. Specifically, we investigate the performance of DKC across various temperature and interaction regimes, providing insights into its feasibility and scalability for molecular systems. A successful experimental implementation of such a collimation would open a range of potential applications especially since the recent report of the first BEC of dipolar molecules \cite{bigagliObservationBoseEinstein2024}. Notably, improved collimation could enable longer interrogation times in experiments using ultracold molecular ensembles\;\cite{D’Incao_2023} or collimated molecular beams\;\cite{Beams_2021}, paving the way for the observation of universal molecular dynamics and for molecular interferometry over extended durations. Molecular DKC could also provide an improved foundation for the preparation of more complex molecules, including Efimov triplets\;\cite{Klauss2017, Xie2020} and larger multiplets\;\cite{Stecher2009, Bazak_2020}. Last but not least, it could allow the collimation of dual-species gases, eliminating the need for multiple lensing techniques\;\cite{Corgier_2020}, and thereby facilitate differential atom interferometry experiments aimed at testing the universality of free fall, a fundamental principle of Einstein's general theory of relativity\;\cite{Will2014, Overstreet2018, Elliott2023, Barrett2022}.

The remainder of this paper is organized as follows. In Section\;\ref{sec:Model}, we present our theoretical framework, key assumptions, and the trapping potential used to model the DKC process. Section\;\ref{sec:Results} evaluates the performance of molecular DKC across condensed, hydrodynamic, and thermal regimes, highlighting key scaling behaviors and interaction effects. Finally, in Section\;\ref{sec:conclu}, we summarize our findings, discuss their implications for experimental realizations, and outline future research directions.

\section{Theoretical model}
\label{sec:Model}

\subsection{Molecular system}

While the results presented here are applicable to a wide range of diatomic molecules, this study specifically examines bosonic Feshbach molecules formed from $^{41}$K and $^{87}$Rb atoms. These atoms are particularly well-suited for many of the previously mentioned dual-species gas studies due to their broad heteronuclear Feshbach resonances and the existence of weakly bound $^{41}$K$^{87}$Rb states at magnetic bias fields below 100 Gauss.

Before molecular association, both atomic species are assumed to be independently cooled and trapped. Although microgravity can enhance the efficiency of this association\;\cite{DIncao2017, Waiblinger2021}, successful ground-based formation of heteronuclear Feshbach molecules has been demonstrated for multiple species, including NaCs, NaK, NaRb, and KRb\;\cite{Voges2020, bigagliObservationBoseEinstein2024, Weber2008, Aikawa2010, Gersema2021}.

We assume that the Feshbach molecules are sufficiently large to neglect couplings between \mbox{$\ell=0$} and higher partial waves due to magnetic and induced dipole-dipole interactions. Furthermore, the molecules are assumed to be initially sufficiently cold that the effects of higher partial waves in molecule-molecule interactions are also neglected. In this regime, the resulting molecular ensemble interacts exclusively via $s$-wave (\mbox{$\ell=0$}) interactions.

While one-photon rotational transitions are suppressed when the trapping light is far-detuned from all excited levels, two-photon transitions tend to be more probable. In a dipolar trap generated by a laser beam with well-defined polarization, rotational excitations will not be driven, as each two-photon stimulated emission releases the same amount of angular momentum that it gained during the initial absorption. In addition, even if polarizations are mixed through some mechanism, rotational excitations remain constrained by parity laws. Finally, during the DKC process, the optical potential is applied very briefly to the molecular ensemble so Fourier broadening defines the minimum spectral width of the applied potential. If this width becomes broad enough for the two-photon detuning to match a rotational excitation, parasitic rotational excitations might happen. However, rotational transitions are typically in the microwave regime, meaning that the two-photon detuning would be far too large for such effects to occur efficiently. What remains is the coupled dynamics of the vibrational and translational motions, which is addressed in Section~\ref{sec:coupl}.

In contrast to bosonic Feshbach molecules formed with fermionic atoms \cite{zirbel2008prl}, $^{41}$K$^{87}$Rb molecules suffer from inelastic loss processes due to both atom-molecule and molecule-molecules collisions \cite{dincao2018JPB}. To avoid atom-molecule collisions, it is essential to purify the molecular sample by removing all residual atoms, as losses from such collisions are significantly faster than those from molecule-molecule collisions due to the lower molecular densities, as demonstrated for bosonic NaRb molecules in Ref.\;\cite{WangNJP2015}. Purification of the sample can be achieved through several methods, including resonant pulses (particularly if the molecular BEC has to be prepared in a lower vibrational state), Stern-Gerlach type separation (taking advantage of the distinct magnetic moments of K, Rb and KRb), or other techniques tailored to the system. During a time-of-flight (TOF) sequence applied to the molecular ensemble, relaxation into lower vibrational states may still occur due to molecule-molecule collisions. This relaxation imposes a time scale on the TOF duration, which depends on the molecular density but also the specific molecular species. Note however that long TOF durations and ultra-cold low-density gases can be achieved in microgravity environments\;\cite{DIncao2017, Waiblinger2021}.

\subsection{Trapping Potential}
\label{sec::Traping_Potential}

We consider an optical dipole trap created by a non-uniform intensity distribution denoted as $\mathcal{I}$. Within the trapping region, we assume that the spatial variation of this intensity is harmonic and isotropic. Experimentally, such a harmonic intensity distribution can be achieved over distances on the order of a millimeter using time-averaged (painted) potentials\;\cite{Albers2022, Herbst24, PhysRevResearch.6.013139}. The potential experienced by the molecules is represented by $V_\mathrm{Mol}(\mathbf{R},\mathbf{r})$, where $\mathbf{R} = ( m_\mathrm{K}\,\mathbf{r}_\mathrm{K} + m_\mathrm{Rb}\,\mathbf{r}_\mathrm{Rb} ) / ( m_\mathrm{K} + m_\mathrm{Rb} )$ denotes its center-of-mass position, and $\mathbf{r} = \mathbf{r}_\mathrm{Rb} - \mathbf{r}_\mathrm{K}$ represents the internuclear coordinate. $\mathbf{r}_\mathrm{K}$ and $\mathbf{r}_\mathrm{Rb}$ are the atomic coordinates, and $m_\mathrm{K}$ and $m_\mathrm{Rb}$ are the atomic masses of K and Rb, respectively. Assuming an atomic approach, which is accurate at large internuclear distances\,\cite{Vexiau_2011}, we write
\begin{equation}
V_\mathrm{Mol}(\mathbf{R},\mathbf{r}) = V_{\mathrm{int}}(r) + \alpha_\mathrm{K} \, \mathcal{I}(\mathbf{r}_\mathrm{K}) + \alpha_\mathrm{Rb} \, \mathcal{I}(\mathbf{r}_\mathrm{Rb}),
\label{eq:Vmol}
\end{equation}
where $V_{\mathrm{int}}(r)$ is the interaction energy between the two atoms, and $\alpha_\mathrm{K}$ and $\alpha_\mathrm{Rb}$ are the dynamic polarizabilities of K and Rb, respectively. If the variation of $\mathcal{I}$ over the spatial extent of the molecule is negligible, then $\mathcal{I}(\mathbf{r}_\mathrm{Rb}) \simeq \mathcal{I}(\mathbf{R}) \simeq \mathcal{I}(\mathbf{r}_\mathrm{K})$, which allows us to approximate the molecular potential as
\begin{equation}
V_\mathrm{Mol}(\mathbf{R},\mathbf{r}) \simeq V_{\mathrm{int}}(r) + \big[\alpha_\mathrm{K} + \alpha_\mathrm{Rb} \big]\,\mathcal{I}(\mathbf{R})\,.
\label{eq:Vmol3}
\end{equation}
This approximation naturally gives rise to the emergence of the molecular polarizability, written as
\begin{equation}
\alpha_\mathrm{Mol} = \alpha_\mathrm{K} + \alpha_\mathrm{Rb}\,.
\end{equation}
It should also be noted that in this approximation, the molecular potential becomes separable in the $(\mathbf{R},\mathbf{r})$ coordinate system. In the harmonic and isotropic case considered here we can write
\begin{equation}
V_\mathrm{Mol}(\mathbf{R},\mathbf{r}) \simeq V_{\mathrm{int}}(r) + \frac{1}{2} M \omega_\mathrm{Mol}^2 R^2\,,
\label{eq:Vmol4}
\end{equation}
where $M = m_\mathrm{K}+m_\mathrm{Rb}$ is the total molecular mass. This approximation assumes that the intensity remains constant over a distance on the order of the molecular size. This condition imposes a constraint on the characteristic size $a_\mathrm{Mol}$ associated with the harmonic trap, given by
\begin{equation}
a_\mathrm{Mol} = \sqrt{\frac{\hbar}{M\omega_\mathrm{Mol}}} \gg a
\label{eq:condition}
\end{equation}
where $a$, the scattering length between the two atoms, is taken as an upper bound for the typical size of the Feshbach molecule \cite{Grimm2006}.

We now set aside the approximation introduced in Eq.\;(\ref{eq:Vmol3}) and revert to the exact expression for the molecular potential given in Eq.\;(\ref{eq:Vmol}). For the harmonic case considered here, the potential takes the form
\begin{equation}
V_\mathrm{Mol}(\mathbf{R},\mathbf{r}) = V_{\mathrm{int}}(r) + \frac{1}{2} \Big[ m_\mathrm{K} \, \omega_\mathrm{K}^2 \, \mathbf{r}_\mathrm{K}^2 + m_\mathrm{Rb} \, \omega_\mathrm{Rb}^2 \, \mathbf{r}_\mathrm{Rb}^2\Big]\,,
\label{eq:Vmol2}
\end{equation}
where the second term (the trapping potential) can be conveniently rewritten as
\begin{equation}
V_\mathrm{Mol}^{(t)}(\mathbf{R},\mathbf{r}) = V_\mathrm{CM}(R) + V_\mathrm{rel}(r) + V_\mathrm{coup}(\mathbf{R},\mathbf{r}) \label{eq:full_pot}
\end{equation}
where the different terms are defined as follows
\begin{subequations}
\begin{eqnarray}
V_\mathrm{CM}(R) & = & \frac{1}{2} \, M \, \omega_\mathrm{Mol}^2 \, R^2\\[0.2cm]
V_\mathrm{rel}(r) & = & \frac{1}{2} \, \mu \, \omega_\mathrm{r}^2 \, r^2\\[0.2cm]
V_\mathrm{coup}(\mathbf{R},\mathbf{r}) & = & - \mu \, \omega_\mathrm{c}^2 \, \mathbf{R} \cdot \mathbf{r} \label{eq:Vcoup}
\end{eqnarray}
\end{subequations}
The reduced mass is given by $\mu = m_\mathrm{K}m_\mathrm{Rb}/M$ and the harmonic frequencies $\omega_\mathrm{Mol}$, $\omega_\mathrm{r}$ and $\omega_\mathrm{c}$ are defined by 
\begin{subequations}
\begin{eqnarray}
M\,\omega_\mathrm{Mol}^2 & = & m_\mathrm{K} \, \omega_\mathrm{K}^2 + m_\mathrm{Rb} \, \omega_\mathrm{Rb}^2\,,\\[0.2cm]
M\,\omega_\mathrm{r}^2 & = & m_\mathrm{Rb} \, \omega_\mathrm{K}^2 + m_\mathrm{K} \, \omega_\mathrm{Rb}^2\,, \label{eq:wrel}\\[0.2cm]
\omega_\mathrm{c}^2 & = & \omega_\mathrm{K}^2 - \omega_\mathrm{Rb}^2\,. \label{eq:wcoup}
\end{eqnarray}
\end{subequations}

From the preceding expressions\;(\ref{eq:Vmol2}) and (\ref{eq:full_pot}), it can be seen that the total molecular potential becomes separable, with $V_\mathrm{Mol}(R,r) \simeq V_\mathrm{CM}(R) + [V_{\mathrm{int}}(r)+V_\mathrm{rel}(r)]$, when the coupling $V_\mathrm{coup}(\mathbf{R},\mathbf{r})$ between center-of-mass and relative motions is negligible. Under this condition, the two degrees of freedom can be treated independently. To determine whether this condition is satisfied, we now express the frequencies $\omega_\mathrm{r}$ and $\omega_\mathrm{c}$ in terms of the center-of-mass frequency $\omega_\mathrm{Mol}$, which is used as the input parameter for the simulations.

With the present harmonic and isotropic potential, we obtain from Eqs.\;(\ref{eq:Vmol}) and\;(\ref{eq:Vmol2}) \mbox{$\alpha_\mathrm{K}\nabla^2\mathcal{I} = 3m_\mathrm{K}\omega_\mathrm{K}^2$} and \mbox{$\alpha_\mathrm{Rb}\nabla^2\mathcal{I} = 3m_\mathrm{Rb}\omega_\mathrm{Rb}^2$}. In the limit where the condition\;(\ref{eq:condition}) is satisfied, we also derive from Eqs.\;(\ref{eq:Vmol3}) and (\ref{eq:Vmol4}) that \mbox{$\alpha_\mathrm{Mol}\nabla^2\mathcal{I} = 3M\omega_\mathrm{Mol}^2$}. Consequently, we can write
\begin{equation}
\frac{m_\mathrm{K}}{\alpha_\mathrm{K}} \, \omega_\mathrm{K}^2 = \frac{m_\mathrm{Rb}}{\alpha_\mathrm{Rb}} \, \omega_\mathrm{Rb}^2 =
\frac{M}{\alpha_\mathrm{K} + \alpha_\mathrm{Rb}}\,\omega_\mathrm{Mol}^2\,.
\end{equation}
Defining the ratio between the two dynamic polarizabilities of the atoms as $p = \alpha_\mathrm{Rb}/\alpha_\mathrm{K}$, we can finally express the harmonic trapping frequencies $\omega_\mathrm{r}$ and $\omega_\mathrm{c}$ as
\begin{subequations}
\begin{eqnarray}
\omega_\mathrm{r} & = & \left[\frac{m_\mathrm{Rb}^2 + p \, m_\mathrm{K}^2}{(p+1) \, m_\mathrm{K} \, m_\mathrm{Rb}}\right]^{\frac{1}{2}}\omega_\mathrm{Mol} \label{eq:omega_r}\\[0.2cm]
\omega_\mathrm{c} & = &  \left[\frac{m_\mathrm{Rb} - p \, m_\mathrm{K}}{(p+1) \, \mu}\right]^{\frac{1}{2}}\omega_\mathrm{Mol} \label{eq:omega_c}
\end{eqnarray}
\end{subequations}
Since the ratio $p$ of the dynamic polarizabilities is fixed at a given wavelength, it follows that selecting one of the frequencies $\omega_\mathrm{Mol}$, $\omega_\mathrm{r}$ or $\omega_\mathrm{c}$ automatically determines all the others.

\subsection{Coupled Dynamics of Molecular Motions}
\label{sec:coupl}

In the following, two key hypotheses will be made: the separability of the vibrational motion from the overall translational motion, and the stability of the Feshbach molecule. To support these hypotheses, it is first necessary to establish a model for estimating both the extent of coupling between these two degrees of freedom and the energy transferred to the vibrational degree of freedom during a typical DKC process. To this end, and to maintain the simplicity of our approach, we adopt a classical model to describe the coupled nuclear motions.

It is important to note that the interaction potential between the atoms, $V_\mathrm{int}(r)$, induces fast oscillatory motion compared to the slower oscillatory dynamics of the trapping potential. For instance, in a Feshbach molecule with a binding energy $E_b = \hbar^2/(2\mu a^2)$ for a scattering length $a=1\,000$\,a.u, the rapid classical period $T_r$ of oscillation in the potential well $V_\mathrm{int}(r)$ can be estimated as
\begin{equation}
T_r = \sqrt{2\mu} \int_{r_\mathrm{min}}^{r_\mathrm{max}} \frac{dr}{\sqrt{-E_b-V_\mathrm{int}(r)}}\,,
\end{equation}
which is typically on the order of a fraction of a microsecond\;\cite{Jasik2023}. By contrast, the trapping frequencies, typically around a hundred Hz, correspond to slow oscillatory periods that are approximately $10^4$ times larger. To address the long-time trapping dynamics, we thus employ the adiabatic approximation, replacing the interaction potential $V_\mathrm{int}(r)$ in the system Hamiltonian with its mean value over a fast oscillatory period, $\langle V_\mathrm{int}\rangle$. Assuming zero angular momentum, the system Hamiltonian becomes
\begin{eqnarray}
H(r,p_r,R,p_R) & = & \frac{p_r^2}{2\mu} + V_\mathrm{rel}(r) + \langle V_\mathrm{int}\rangle + \frac{p_R^2}{2M} \nonumber \\
               &   &  + V_\mathrm{CM}(R) + V_\mathrm{coup}(\mathbf{R},\mathbf{r})\,.
\end{eqnarray}
The coupling potential $V_\mathrm{coup}(\mathbf{R},\mathbf{r})$, as defined in Eq.\;(\ref{eq:Vcoup}), is proportional to $\mathbf{R} \cdot \mathbf{r} = r \, R \, \cos{(\theta)}$, where $\theta$ is the angle between the two coordinates. The coupling is therefore maximum when $\mathbf{R}$ and $\mathbf{r}$ are collinear, \emph{i.e.} when $\theta = n \pi$. In the following, we will consider the worst-case scenario of maximum coupling, given by
\begin{equation}
V_\mathrm{coup}(R,r) = - \, \mu \, \omega_\mathrm{c}^2 \, R \, r.
\end{equation}
Using this expression, the classical Hamiltonian equations of motion for both coordinates can be written as follows
\begin{subequations}
\begin{eqnarray}
\label{eq:osc_R}
\Ddot{R} + \omega_\mathrm{Mol}^2 \, R & = & (\mu / M) \, \omega_\mathrm{c}^2 \, r\,, \\[0.1cm]
\label{eq:osc_r}
\Ddot{r} + \omega_\mathrm{r}^2 \, r   & = & \omega_\mathrm{c}^2 \, R\,,
\end{eqnarray}
\end{subequations}
and the eigenfrequencies associated with the coupled system are given by
\begin{equation}
2\omega_\pm^2 =
\omega_\mathrm{Mol}^2 + \omega_\mathrm{r}^2
\pm \sqrt{(\omega_\mathrm{Mol}^2 - \omega_\mathrm{r}^2)^2 + 4(\mu/M)\omega_\mathrm{c}^4}\,.
\label{eq:normalmodes}
\end{equation}
The normal modes corresponding to these eigenfrequencies $\omega_\pm$ can be expressed as
\begin{eqnarray}
\label{eq:zp}
z_+ & = & R - r \cdot \gamma/(1+\gamma)\,,\\
\label{eq:zm}
z_- & = & R + r \cdot 1/(1+\gamma)\,,
\end{eqnarray}
where $\gamma$ denotes the mass ratio $\gamma= m_\mathrm{Rb}/m_\mathrm{K}$. It is worth noting that this change of coordinates is independent of the frequencies $\omega_\textrm{Mol}$, $\omega_\textrm{r}$ and $\omega_\textrm{c}$. To follow the coupled dynamics during the DKC process, we implement linear on/off ramps for the harmonic trap, described by
\begin{equation}
\omega_\mathrm{Mol}^2(t) = \left\{
\begin{array}{ll}
\omega_0^2 \cdot t/t_\mathrm{r} &
\mathrm{for}\; 0 \leqslant t < t_\mathrm{r}  \\[0.1cm]
\omega_0^2                          &
\mathrm{for}\; t_\mathrm{r} \leqslant t \leqslant t_\mathrm{DKC} - t_\mathrm{r}  \\[0.1cm]
\omega_0^2 \cdot (t_\mathrm{DKC}-t)/t_\mathrm{r} &
\mathrm{for}\; t_\mathrm{DKC}\!-\!t_\mathrm{r}\! < t \leqslant t_\mathrm{DKC}
\end{array}
\right.\nonumber
\label{eq:DKCramp}
\end{equation}
where $t_\mathrm{r}$ is the duration of the on/off ramps, and $t_\mathrm{DKC}$ is the total duration of the DKC. Here, $\omega_0$ represents the maximum trapping frequency achieved during the DKC. From Eqs.\;(\ref{eq:omega_r}) and (\ref{eq:omega_c}), it follows that the frequencies $\omega_\mathrm{r}$ and $\omega_\mathrm{c}$ become themselves time-dependent. In the new system of coordinates Eqs.\;(\ref{eq:osc_R}) and (\ref{eq:osc_r}) become
\begin{subequations}
\begin{eqnarray}
\label{eq:osc_zp}
\Ddot{z}_+ + \omega_+^2(t) \, z_+ & = & 0\\[0.1cm]
\Ddot{z}_- + \omega_-^2(t) \, z_- & = & 0
\label{eq:osc_zm}
\end{eqnarray}
\end{subequations}
and are fully uncoupled. From Eqs.\;(\ref{eq:omega_r}), (\ref{eq:omega_c}) and (\ref{eq:normalmodes}), it follows that $\omega_\pm^2(t)=\alpha_\pm\,\,\omega_\mathrm{Mol}^2(t)$ where \mbox{$\alpha_+ = (1+\gamma)/(1+p)$} and \mbox{$\alpha_{-} = p\,\alpha_+/\gamma$}. Consequently, the squared frequencies $\omega_\pm^2(t)$ remain constant for \mbox{$t_\mathrm{r} \leqslant t \leqslant t_\mathrm{DKC} - t_\mathrm{r}$} and vary linearly with time during the on/off ramps. The solutions $z_{+}(t)$ and $z_{-}(t)$ of Eqs.\;(\ref{eq:osc_zp}) and (\ref{eq:osc_zm}) are thus analytical and can be written as linear combinations of the Airy functions of the first and second kind during the switch-on ramp
\begin{equation}
z_\pm(t) = a_\pm\,\mathrm{Ai}\big[-\eta_\pm t\big] +
b_\pm\,\mathrm{Bi}\big[-\eta_\pm t\big]\,,
\end{equation}
as trigonometric functions for $t_\mathrm{r} \leqslant t \leqslant t_\mathrm{DKC} - t_\mathrm{r}$
\begin{equation}
z_\pm(t) = c_\pm \cos\big[\alpha_\pm^{1/2}\omega_0\,t\big] + s_\pm \sin\big[\alpha_\pm^{1/2}\omega_0\,t\big]\,,
\end{equation}
and again as Airy functions during the switch-off ramp
\begin{equation}
z_\pm(t) = a'_\pm\,\mathrm{Ai}\big[\eta_\pm (t-t_\mathrm{DKC})\big] +
b'_\pm\,\mathrm{Bi}\big[\eta_\pm (t-t_\mathrm{DKC})\big]\,.
\end{equation}
Here, the frequencies $\eta_\pm$ are given by \mbox{$\eta_\pm = (\alpha_\pm\,\omega_0^2/t_\mathrm{r})^{1/3}$}. The coefficients $a_\pm$, $b_\pm$, $c_\pm$, $s_\pm$, $a'_\pm$ and $b'_\pm$, are determined by the boundary conditions at times $t=0$, $t=t_\mathrm{r}$ and $t=t_\mathrm{DKC}-t_\mathrm{r}$. By inverting Eqs.\;(\ref{eq:zp}) and (\ref{eq:zm}), we finally obtain the exact solutions $R(t)$ and $r(t)$ for the coupled dynamics described by Eqs.\;(\ref{eq:osc_R}) and (\ref{eq:osc_r}).

\begin{figure}[t!]
\centering
\includegraphics*[width=\columnwidth]{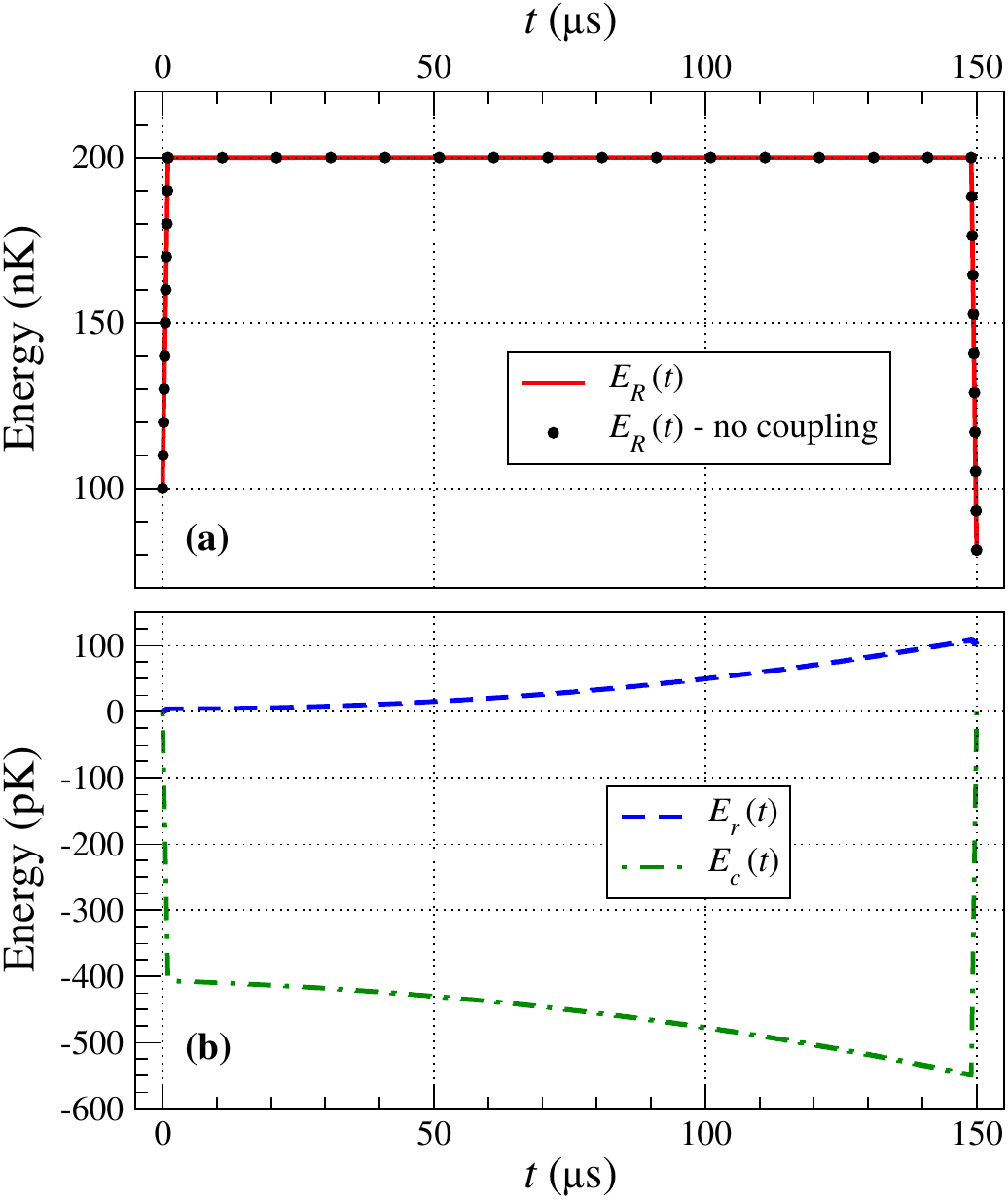}
\caption{Time evolution of the energies $E_\mathrm{R}(t)$ (red solid line and black dots in panel (a)), $E_\mathrm{r}(t)$ (blue dashed line in panel (b)), and $E_\mathrm{c}(t)$ (green dash-dotted line in panel (b)) during a typical DKC process. The solid red line represents results obtained including the coupling between the center-of-mass and vibrational motions, while the black dots correspond to calculations where this coupling is neglected.}
\label{fig:clas_dyn}
\end{figure}

Figure\;\ref{fig:clas_dyn} shows the temporal evolution of the energies associated with the center-of-mass motion
\begin{equation}
E_\mathrm{R}(t) = \frac{1}{2} M \Big[ \omega_\mathrm{Mol}^2(t)\,R^2(t) + \dot{R}^2(t) \Big]\,,
\label{eq:comEn}
\end{equation}
the molecular vibration
\begin{equation}
E_\mathrm{r}(t) = \frac{1}{2} \mu \Big[\omega_\mathrm{r}^2(t)\,R^2(t) + \dot{r}^2(t) \Big]\,,
\label{eq:vibEn}
\end{equation}
and the coupling term
\begin{equation}
E_\mathrm{c}(t) = - \mu\,\omega_\mathrm{c}^2(t)\,R(t)\,r(t)
\label{eq:coupEn}
\end{equation}
for a typical DKC process. These results are based on the assumption that the optical dipole trap operates at a wavelength of 
$\lambda = 2\,000$\,nm which, as discussed below, is actually a worst-case scenario. At this wavelength, the ratio $p$ of the dynamic polarizabilities of Rb and K is approximately $p = \alpha_\mathrm{Rb} / \alpha_\mathrm{K} \simeq 1.10$\;\cite{alpha_K,alpha_Rb}. The total duration of the DKC is $t_\mathrm{DKC}=150\,\mu$s, including the switch-on and switch-off ramps, each of them lasting 1\,$\mu$s. The maximum trapping frequency is set to \mbox{$\omega_0 = 2 \pi \times 100$\,Hz}. The initial conditions are chosen as follows. For the $R$ coordinate, corresponding to the center-of-mass motion, the total initial energy is set to 200\,nK, a value chosen to be within the typical order of magnitude for Bose-Einstein condensate energies. This energy is equally partitioned between its kinetic and potential components, each contributing 100\,nK. Specifically, this corresponds to \mbox{$R(0)=4.06\,\mu$m} and \mbox{$\dot{R}(0)=2.55\,\mu$m/ms}. For the vibrational coordinate $r$, we consider a very weakly bound molecule formed via Feshbach resonance. Such molecules are likely to be at very large internuclear distances, near or beyond the outer turning point of the molecular potential. The initial condition for this coordinate is set to $r(0) = 1\,000$\,a.u., with an initial velocity of $\dot{r}(0) = 0$.

In part (a) of Fig.\;\ref{fig:clas_dyn}, the solid red line shows the time evolution of the center-of-mass energy $E_\mathrm{R}(t)$ [Eq.\,(\ref{eq:comEn})] in a calculation that includes the coupling with the vibrational motion. At the initial time $t=0$, this energy is 100\,nK, as the harmonic potential is off at this point. By the end of the switch-on ramp, after 1\,$\mu$s, the total energy reaches the expected value of 200\,nK. Subsequently, this energy remains nearly constant throughout the collimation process, before decreasing during the switch-off ramp to a final value of approximately 81\,nK. The black dots on the same graph represent the results of a similar calculation in which the coupling with the vibrational degree of freedom has been neglected. The results of this second simulation are superimposed on those obtained when the coupling is included. At $t=t_\mathrm{DKC}=150\,\mu$s, the relative difference between the two final energies is only $0.045\%$. We can conclude that, with respect to the center-of-mass dynamics, the coupling with the vibrational degree of freedom can be safely neglected. It is therefore justified to disregard this coupling when describing the DKC process of the molecular cloud.

In part (b) of Fig.\;\ref{fig:clas_dyn}, the blue dashed line represents the evolution of the vibrational energy $E_\mathrm{r}(t)$ [Eq.\,(\ref{eq:vibEn})], while the green dash-dotted line shows the evolution of the coupling energy $E_\mathrm{c}(t)$ [Eq.\,(\ref{eq:coupEn})] between the two degrees of freedom in the calculation which includes this coupling. This graph reveals that the coupling energy $E_\mathrm{c}(t)$ ranges between $-400$ and $-550$\,pK, values that are several hundred times smaller than the center-of-mass energy $E_\mathrm{R}(t)$. This substantial energy gap explains why the coupling has minimal influence on the center-of-mass dynamics. This calculation also shows that during the DKC process, the energy gain in the vibrational coordinate is about 100\,pK. This value is significantly smaller than the binding energy of the Feshbach molecule, which can be estimated as $E_b = \hbar^2/(2\mu a^2) \simeq 6\,\mu$K for a scattering length of $a=1\,000$\,a.u. This energy gain is approximately $6 \times 10^4$ times smaller than the binding energy of the molecule, making it highly unlikely for the molecule to dissociate during the DKC process.

It is important to note that these results were obtained under the assumption that the dipolar optical trap operates at a wavelength of $\lambda = 2\,000$\,nm. This could raise questions about the validity of our conclusions regarding the decoupling of the center-of-mass motion and the molecular stability at shorter wavelengths, particularly since wavelengths between 800 and 1\,100\,nm are commonly used for atomic and molecular trapping. Refs.\;\cite{alpha_K,alpha_Rb} illustrate the evolution of the dynamic polarizabilities of Rb and K as a function of wavelength. From these references, it can be concluded that the ratio $p$ of these atomic polarizabilities increases monotonically as the wavelength decreases from 2\,000\,nm to 800\,nm. In addition, according to Eq.\;(\ref{eq:omega_c}), the coupling term decreases as $p$ increases. This trend indicates that the decoupling of the center-of-mass motion and the molecular stability will be further enhanced at shorter wavelengths commonly used in experimental setups, such as those around 1\,064\,nm.

Finally, from Eq.\;(\ref{eq:omega_c}) we can see that the coupling term vanishes when $p = p_\mathrm{opt} = m_\mathrm{Rb} / m_\mathrm{K}$. In this case Eq.\;(\ref{eq:omega_r}) gives $\omega_\mathrm{r} = \omega_\mathrm{Mol}$ and since there is no coupling between the two harmonic oscillators for this particular ratio $p_\mathrm{opt}$ of polarizabilities, the vibrational and translational coordinates oscillate independently at the same frequency $\omega_\mathrm{Mol}$. Numerically we get $p_\mathrm{opt} \simeq 2.12$, and based on \cite{alpha_K,alpha_Rb}, this particular ratio of dynamic polarizabilities corresponds to the laser wavelength \mbox{$\lambda_\mathrm{opt} \simeq 808.25$\,nm}. For an optical trap operating at this so-called `magic' wavelength, there is a complete decoupling between the center-of-mass motion and the relative motion of the atoms.

In summary, our findings confirm a strong decoupling between the vibrational and translational degrees of freedom across a broad range of trapping wavelengths. Additionally, three key parameters govern the stability of the molecule during the delta-kick process. The first one is the coupling strength $\omega_c^2$ seen in Eqs.\,(\ref{eq:osc_R}-\ref{eq:osc_r}). This parameter depends on the relative polarizabilities of the atomic species forming the molecule and can be tuned by adjusting the trapping wavelength. However, complete suppression of this coupling occurs only at a specific wavelength. For commercially available lasers and the molecule considered in this work, the coupling strength remains non-zero. The second parameter is the initial center-of-mass energy. Indeed, since the coupling is non-zero, the risk of molecular dissociation increases with the initial center-of-mass energy. Given that we are dealing with Feshbach molecules, which have a low binding energy, an initial center-of-mass energy of this magnitude combined with non-zero coupling, could significantly compromise the molecule's stability. Finally, the last parameter is the duration of the delta-kick. In our chosen scenario, energy transfer is inefficient because the kick duration (150\,$\mu$s) is much shorter than the characteristic timescale required for significant energy transfer between the center-of-mass and the vibrational degrees of freedom. This coupling timescale, on the order of $1/\omega_c$, corresponds to a few milliseconds. Since the delta-kick operates on a far shorter timescale, energy transfer remains negligible, minimizing the impact of coupling and ensuring molecular stability.

\section{DKC performance in the different regimes}
\label{sec:Results}

We consider molecular ensembles in both the thermal and BEC regimes. Taking advantage of the quasi-separability of the center-of-mass and relative coordinates, we focus on the DKC dynamics of the cloud with respect to the translational coordinate $R$, while factorizing out the vibrational dynamics. The molecular ensemble is initially trapped in a dipolar potential\;\cite{GRIMM200095, Roy2016, AlbersThesis, PhysRevResearch.6.013139}. The DKC sequence, as originally described in \cite{PhysRevLett.78.2088}, begins with a free expansion phase of duration $t_\mathrm{preTOF}$, followed by a brief flashing of the trap for $t_\mathrm{DKC}$, and concludes with a final TOF phase of duration $t_\mathrm{TOF}$. Finally, if needed, the molecules can be dissociated\;\cite{PhysRevA.95.012701} into two ultra-cold atomic ensembles with reduced expansion energies compared to the initial atomic ensembles.

\subsection{Scaling Dynamics Across Condensed and Thermal Regimes}

To evaluate the dynamics of the molecular ensemble, we use scaling laws of the form $\sigma(t) = \sigma(0)\,\lambda(t)$, where $\sigma(0)$ represents the initial size of the ensemble, and the equation of motion concerns only the scaling factor $\lambda(t)$.

In the condensed regime, we consider two different scaling approaches. The first approach is based on the Thomas-Fermi approximation, as described in \cite{PhysRevLett.77.5315, PhysRevA.55.R18}, which neglects the kinetic energy term in the Gross-Pitaevskii equation \cite{Pethick02}. The scaling equation for this approach is given by
\begin{equation}
\ddot{\lambda}(t) + \omega_{\mathrm{Mol}}^2(t)\,\lambda(t) = \frac{\omega_{\mathrm{Mol}}^2(0)}{\lambda^4(t)}\,.
\label{eq:TF_scaling}
\end{equation}
For this model to remain valid, the kinetic term in the Gross-Pitaevskii equation must be negligible. This condition requires either a large number of atoms or a large dimer-dimer scattering length $a_\mathrm{dd}$, which implies strong interactions in both cases. If neither of these conditions is met, this approach is no longer applicable, and one must rely on a second approach that uses a variational method\;\cite{Perez96PRL, Perez97PRA}. This method includes a correction term accounting for the kinetic energy. The scaling equation for this approach is given by
\begin{equation}
\ddot{\lambda}(t) + \omega_{\mathrm{Mol}}^2(t)\,\lambda(t) =
\frac{\omega_{\mathrm{Mol}}^2(0)}{\lambda^4(t)} +
\alpha \frac{ \omega_{\mathrm{Mol}}^2(0)}{\lambda^3(t)}\,,
\label{eq:var_scaling}
\end{equation}
where $\alpha = (\pi/2)^\frac{2}{5}[a_\mathrm{Mol}/(N a_\mathrm{dd})]^\frac{4}{5}$ and the expression of $a_\mathrm{Mol}$ is given in Eq.\;(\ref{eq:condition}). This correction term vanishes in the strong interaction limit, reducing the model to the Thomas-Fermi approximation. In both approaches, the initial size of the ensemble is set to the Thomas-Fermi radius
\begin{equation}
\sigma(0) = a_\mathrm{Mol}
\left(\frac{15\,Na_\mathrm{dd}}{a_\mathrm{Mol}}\right)^{\!\frac{1}{5}}\,.
\label{eq:init_size_TF}
\end{equation}

For both the hydrodynamic and thermal regimes, the previous equations cannot be applied, as they are only valid for pure condensates. Instead, we adopt the approach developed in \cite{PhysRevA.66.033613, PhysRevA.68.043608} and adapt it to our isotropic case. This leads to the following scaling equation
\begin{equation}
\ddot{\lambda}(t) + \omega_{\mathrm{Mol}}^2(t)\,\lambda(t) =
(1 - \xi)\;\frac{\omega_{\mathrm{Mol}}^2(0)}{\lambda^3(t)}
+ \xi\;\frac{\omega_{\mathrm{Mol}}^2(0)}{\lambda^4(t)}\,,
\label{eq:hydro}
\end{equation}
where $E_\mathrm{mf} = 4 \pi\hbar^2a_\mathrm{dd}n_0/(\sqrt{2} \, M)$ and $\xi = E_\mathrm{mf}/(E_{\mathrm{mf}}+k_B T)$. Here, $T$ is the temperature of the system, and $n_0$ is the initial average density. It is straightforward to verify that in the limit $k_B T \ll E_\mathrm{mf}$, this equation reduces to the Thomas-Fermi scaling equation once again. In the hydrodynamic regime, the initial size $\sigma(0)$ corresponds to the size of a thermal cloud at temperature $T$, given by
\begin{equation}
\sigma(0) = \sqrt{\frac{k_B T}{m}} \, \omega_\mathrm{Mol} \, \sqrt{1-\xi}\,.
\label{eq:init_size_hydro}
\end{equation}

\subsection{Performance Analysis in Condensed Regimes}

To evaluate the performance of the collimation across the different regimes, we compare the expansion energy gain, defined as $\mathcal{G} = E_i/E_f$, where $E_i$ is the expansion energy without collimation, and $E_f$ is the expansion energy after collimation. We express the expansion energy in units of temperature, using $k_B T = m\dot{\sigma}^2$, where $\dot{\sigma}$ denotes the time derivative of $\sigma$.

To simplify the comparison between each case, we use a constant time-of-flight (TOF) duration of 14.9\,ms before the kick. After this TOF, the cloud size is approximately \mbox{$\sigma(t_\mathrm{preTOF}) \approx 20$\,$\mu$m} in the case of a condensed cloud. Such a small size ensures that the harmonic approximation remains completely valid, as the entire BEC is confined to the center of the trap.

\begin{figure}[t!]
\centering
\includegraphics*[width=\columnwidth]{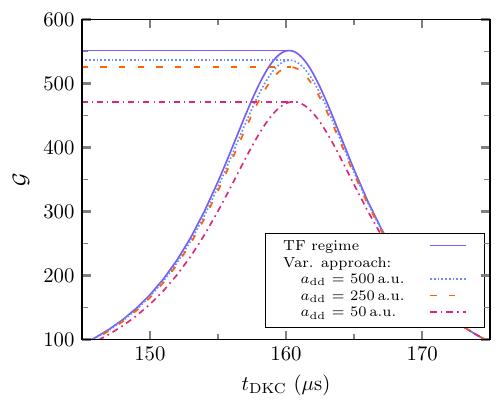}
\caption{Expansion energy gain $\mathcal{G} = E_i/E_f$ as a function of the DKC duration for both the Thomas-Fermi regime (purple solid line) and the variational approach at different scattering lengths. While the gain does not depend on the scattering length in the Thomas-Fermi approach, the variational approach shows variations as the scattering length decreases from 500\,a.u. (blue dotted line) to 250\,a.u. (orange dashed line) and finally to 50\,a.u. (red dash-dotted line). As expected, higher interaction strengths bring the gain of the variational method closer to that of the Thomas-Fermi regime.}
\label{fig::gain_TF_var}
\end{figure}

Starting with the pure BEC case, we first consider the Thomas-Fermi regime. For $5 \times 10^4$ molecules trapped in a $2\pi \times 100$\,Hz isotropic trap, the expansion energy before the kick is 25.2\,nK. The expansion energy gain is shown in Fig.\;\ref{fig::gain_TF_var} as a function of the DKC duration. We achieve a maximum gain slightly above $\mathcal{G} = 550$ in the Thomas-Fermi high-interaction limit (purple solid line), corresponding to an expansion energy below 50\,pK.

For weaker interactions ($a_\mathrm{dd} = 500$\,a.u., $250$\,a.u. and $50$\,a.u., represented by the blue dotted, orange dashed and red dash-dotted lines respectively), the maximum gain consistently exceeds $\mathcal{G} = 450$. This implies that the optimal expansion energy is always below 60\,pK. Additionally, the gain remains above 100 within the range $t_\mathrm{opt} \pm 14 \, \mu$s, providing a sufficiently large time window for experimental realization.

Finally, the larger the scattering length, the closer the results of the variational method approach those of the Thomas-Fermi regime. This behavior can be readily understood from the scaling equations: as the interactions increase, the interaction-dependent correction term becomes negligible, causing the equations to reduce to those of the Thomas-Fermi approximation.

\subsection{Hydrodynamic and Thermal Regime Behavior}

In the hydrodynamic approach, varying the temperature or the interactions produces qualitatively similar effects. Specifically, low temperatures are equivalent to high interactions, while high temperatures correspond to low interactions. Finally, at sufficiently high temperatures, the molecules no longer interact with each other and the scaling equation (\ref{eq:hydro}) reduces to the non-interacting Ermakov equation\;\cite{Ermakov, Pinney, Lewis}
\begin{equation}
\ddot{\lambda}(t) + \omega_{\mathrm{Mol}}^2(t)\,\lambda(t) =
\frac{\omega_{\mathrm{Mol}}^2(0)}{\lambda^3(t)}\,.
\label{eq:Ermakov}
\end{equation}

\begin{figure}[t!]
    \centering
    \includegraphics*[width=\columnwidth]{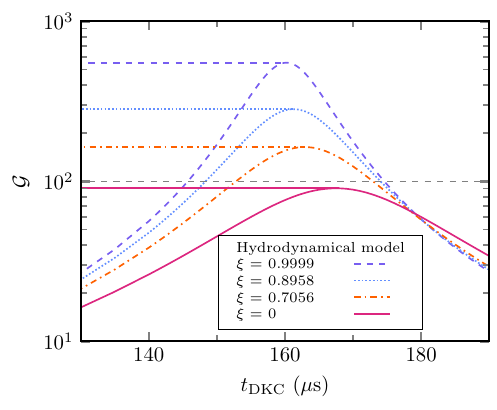}
    \caption{Expansion energy gain $\mathcal{G} = E_i/E_f$ as a function of the DKC duration for a molecular ensemble at finite temperature, with the molecule-molecule scattering length fixed at $a_\mathrm{dd} = 500$\,a.u.. Different values of $\xi$ are investigated, corresponding to various thermal regimes. For $\xi = 0.9999$ (2\,nK, purple dashed line), the system is in the condensed regime. For $\xi = 0.8958$ (30\,nK, blue dotted line) and $\xi = 0.7056$ (50\,nK, orange dash-dotted line) the system transitions to the hydrodynamic regime. Finally, $\xi = 0$ (red solid line) corresponds to a purely thermal ensemble. The maximum gains achieved in each case are as follows: 550 for the condensed regime $(\xi=0.9999)$, 282 for $\xi=0.8958$, 164 for $\xi=0.7056$, and 90 for the thermal regime $\xi=0$.}
    \label{fig::gain_hydro}
\end{figure}

In Fig.\;\ref{fig::gain_hydro} we analyze the gain for four different temperatures with a dipolar scattering length of $a_\mathrm{dd}=500$\,a.u.: 2\,nK (purple dashed line), 30\,nK (blue dotted line), 50\,nK (orange dash-dotted line) and 1\,$\mu$K (red solid line), which correspond to $\xi = 0.9999$, $0.8958$, $0.7056$, and $0$, respectively. The 2\,nK case represents the condensed regime discussed previously, while the 1\,$\mu$K case corresponds to a purely thermal ensemble. The two intermediate temperatures, 30\,nK and 50\,nK, lie within the hydrodynamic regime.

As expected from the scaling Eqs.\;(\ref{eq:TF_scaling}) and (\ref{eq:var_scaling}), the nearly-BEC case (2\,nK, purple dashed line) reaches almost the same maximum gain as the pure BEC case, around $\mathcal{G} = 550$. However, at 2 nK, $\mathcal{G}$ exhibits a slight reduction compared to the pure BEC case, due to the nonzero temperature. The maxima for $\xi = 0.8958$, $0.7056$ and $0$ are 282, 164 and 90, respectively. This decrease correlates with increasing temperature and the associated weaker interactions. 

\subsection{Time Scale Considerations}

These theoretical results, while not surpassing the state-of-the-art collimation achieved experimentally with atoms \cite{PhysRevLett.127.100401}, are of the same order of magnitude. This limitation arises from the conservative value chosen for $t_\mathrm{preTOF}$. According to the Heisenberg uncertainty principle, a larger size at the kick results in a smaller momentum width after the kick. Consequently, with larger values of $t_\mathrm{preTOF}$ it would be possible to achieve even smaller expansion energies. However, this would increase the cloud size at the kick, with molecules further away from the trap center.

Similarly, by keeping $t_\mathrm{preTOF}$ constant but increasing the trap frequency, a larger size at the kick could be achieved, resulting in a smaller final expansion energy. However, the duration of the DKC is governed by $t_\mathrm{DKC} \simeq \big[\sqrt{2\pi}\,\omega_\mathrm{Mol}^2\,t_\mathrm{preTOF}\big]^{-1}$ \cite{PhysRevLett.78.2088}. Therefore, increasing either $t_\mathrm{preTOF}$ or the trap frequency would reduce the collimation time, making the procedure more challenging to implement experimentally.

\section{Summary and Conclusion}
\label{sec:conclu}

In this work, we have investigated the extension of the DKC technique to heteronuclear Feshbach molecules, focusing on its performance across different interaction and temperature regimes. We have demonstrated that this technique, applied to molecular ensembles in both condensed and thermal regimes, can achieve substantial reductions in expansion energy, with theoretical predictions in the tens of picoKelvin range. Specifically, in the condensed regime, our results reveal expansion energy gains exceeding 550, corresponding to expansion energies below 50 picoKelvin. These results highlight the potential of DKC as a robust and efficient tool for controlling ultracold molecular systems.

Furthermore, we have shown that vibrational and translational motions remain strongly decoupled throughout the process, ensuring molecular stability during the delta-kick. This fundamental property reinforces the reliability of DKC for ultracold molecules, as it prevents detrimental energy transfer between internal and external degrees of freedom, which could otherwise reduce collimation efficiency and compromise molecular integrity.

Molecular DKC could be utilized in space experiments, including the upcoming BECCAL apparatus \cite{fryeBoseEinsteinCondensateCold2021}, where the implementation of dipole traps is planned, or in existing experiments that already produce Bose-Einstein condensates of Feshbach molecules \cite{bigagliObservationBoseEinstein2024}. This approach could also be adapted for atom-chip setups that produce mixtures, such as the Cold Atom Laboratory on the International Space Station \cite{Elliott2023}, or for other microgravity facilities \cite{fryeBoseEinsteinCondensateCold2021, Lotz2017, Condon2019, Rudolph2015}. However, in the latter case, care must be taken to ensure that the Feshbach field does not interfere with the trapping potential. Additionally, the magnetic traps generated by such atom-chip setups tend to become anharmonic at larger distances, necessitating verification that the molecules do not reach these anharmonic regions.

Beyond the immediate gains in collimation performance, our findings carry broad implications for precision measurements and quantum technologies. For example, molecular DKC offers a scalable solution for the simultaneous collimation of dual-species ensembles, circumventing the complexities of species-specific lensing. This is particularly relevant for experiments seeking to test fundamental principles such as the universality of free fall (UFF) in dual-species atom interferometry. By reducing systematic uncertainties and enabling extended interrogation times, molecular DKC could significantly enhance the sensitivity and precision of such experiments, offering a new avenue for probing the foundational aspects of general relativity. In addition, our work also sets the stage for future advancements in molecular interferometry.

The ability to collimate molecular ensembles with high precision opens doors to observing universal few-body dynamics, exploring Efimov states, and preparing larger molecular complexes with controlled quantum properties. The scalability of DKC to various molecular systems, coupled with its compatibility with existing trapping and cooling technologies, underscores its versatility for a wide range of applications in quantum physics and ultracold chemistry.

Nevertheless, challenges remain in bridging the gap between theoretical predictions and experimental realizations. The need for fine control over parameters such as the pre-TOF duration and the trap frequency underscores the importance of precise experimental design, as for the collimation of ultracold atomic ensembles. Optimizing these parameters could push the performance of molecular DKC even closer to its theoretical limits.

In summary, our study establishes DKC as a promising tool for advancing the control and manipulation of ultracold molecular systems. Its ability to significantly reduce expansion energy highlights its potential for implementation in experiments requiring long observation times, such as precision tests of fundamental physics\,\cite{Safronova2018}. Furthermore, if the collimated ensemble can be re-trapped using shallow, long-range optical traps such as those generated by painted potentials, it could represent a crucial step toward cooling heteronuclear molecules to quantum degeneracy. As experimental techniques continue to evolve, we anticipate that molecular DKC will play a pivotal role in unlocking new frontiers in both fundamental research and technological applications.

\section*{Acknowledgments}
TE thanks Ali Mouttaki, Stefan Seckmeyer and Pablo Sesma for fruitful discussions. TE acknowledges funding by  the “ADI 2022” project of the IDEX Paris-Saclay, ANR-11-IDEX-0003-02. JPD and JRW are supported by the National Aeronautics and Space Administration through a contract with the Jet Propulsion Laboratory, California Institute of Technology. NG, EMR and TE gratefully acknowledge financial support from the Deutsche Forschungsgemeinschaft (DFG) through SFB 1227 (DQ-mat) within Project A05, Germany’s Excellence Strategy (EXC-2123 QuantumFrontiers Grants No. 390837967), and through the QuantERA 2021 co-funded project No. 499225223 (SQUEIS). NG, EMR and TE also thank the German Space Agency (DLR) for funds provided by the German Federal Ministry for Economic Affairs and Climate Action (BMWK) due to an enactment of the German Bundestag under Grant No. 50WM2450A (QUANTUS-VI), No. 50WM2253A (AI-Quadrat), and No. 50WM2545A (CAL-III).

\bibliographystyle{unsrt}
\bibliography{biblio}

\end{document}